\documentclass{article}
\usepackage{spconf,amsmath,graphicx,amssymb,algorithm,algpseudocode}
\usepackage{float}
\newfloat{pseudocode}{thp}{lop}
\floatname{pseudocode}{Pseudocode}

\title{Unveiling The Tree: \\ A Convex Framework for Sparse Problems}

\name{Tarek A. Lahlou \qquad Alan V. Oppenheim\thanks{The authors wish to thank Analog Devices, Bose Corporation, and Texas Instruments for their support of innovative research at MIT and within the Digital Signal Processing Group.}}
\address{Digital Signal Processing Group \\Massachusetts Institute of Technology }

\begin{document}
\ninept
\maketitle

\begin{abstract}
	This paper presents a general framework for generating greedy algorithms for solving convex constraint satisfaction problems for sparse solutions by mapping the satisfaction problem into one of graph traversal on a rooted tree of unknown topology. For every pre-walk of the tree an initial set of generally dense feasible solutions is processed in such a way that the sparsity of each solution increases with each generation unveiled. The specific computation performed at any particular child node is shown to correspond to an embedding of a polytope into the polytope received from that nodes parent. Several issues related to pre-walk order selection, computational complexity and tractability, and the use of heuristic and/or side information is discussed. An example of a single-path, depth-first algorithm on a tree with randomized vertex reduction and a run-time path selection algorithm is presented in the context of sparse lowpass filter design.
\end{abstract}


\section{Introduction}
\label{sec:intro}
	Sparsity principles, in a broad sense, have been and continue to be associated with desirable properties spanning a wide range of disciplines including sampling theory, model-order reduction, the design of power efficient systems, architectures for deep learning, etc. From the compressive sensing paradigm to the design of sparse filters, solving a convex constraint satisfaction problem for a maximally sparse solution is computationally difficult in general \cite{difficult}. This paper focuses  attention to the class of sparse constraint satisfaction problems which are readily described using a non-convex optimization problem of the form 
	\begin{equation}
		\widehat{\underline{x}} \in \arg \min_{\underline{x}\in\mathcal{S}} \|\underline{x}\|_{0} \label{eq:sparse_problem}
	\end{equation}
	where $\|\cdot\|_{0}$ indicates the number of non-zero coordinates in its argument and $\mathcal{S}$ is a generally convex set. Conventional sparse recovery methods used in compressive sampling and elsewhere can be broadly categorized into one of two classes: convex optimization algorithms and greedy iterative methods. When $\mathcal{S}$ satisfies particular constraint qualification conditions, e.g., the restricted isometry property, Eq.~\ref{eq:sparse_problem} and a convex relaxation of the objective function to the  $1$-norm have been shown to produce identical solutions \cite{CS}. The verification of such qualifications, however, is often itself a computationally expensive or intractable task. 

	Understanding the conditions under which various optimality guarantees can be made as well as the gracefulness with which such guarantees degrade when the conditions are violated but almost satisfied has been the focus of much attention in the literature, especially surrounding the performance of greedy methods. In fact, greedy methods often provably result in a maximally sparse solution for particular classes of problems \cite{greedy}. A well-known strategy involves incrementally decreasing the sparsity of a single solution in accordance with a suitable error metric until a particular stopping criterion is met. For example, (orthogonal) matching pursuit is a method of this type \cite{OMP}. Another strategy, encompassing methods such as Iterative Hard Thresholding (IHT), Subspace Pursuit, and Compressive Sampling Matching Pursuit (CoSaMP), iteratively thresholds dense solutions to a predetermined sparsity level while simultaneously attempting to decrease an error metric \cite{IHT,SP,CoSaMP}. The framework presented in this paper fundamentally differs from these strategies in that the sparsity of many solutions is monotonically increased via tree-based processing while feasibility with respect to Eq.~\ref{eq:sparse_problem} is maintained at every stage. In addition, advanced knowledge or {\it a priori} assumptions on optimal sparsity levels or patterns is not required. 

	Several problems of broad interest to the signal processing community and elsewhere do not meet the constraint qualifications of compressive sensing, an example of which is presented in Section~3 in the context of sparse lowpass filter design. Specialized algorithms that make use of heuristics and/or side information are often designed and lead to sufficiently sparse solutions for these problems. The class of algorithms proposed in this paper does not require such constraint qualifications and is therefore of use when such qualifications either fail or cannot be verified. Further, any side information or heuristic knowledge of the sparsity pattern of an optimal solution is easily incorporated into an algorithm of this framework with only minimal adjustment. 

	In Section~2 the general framework for generating greedy tree-search algorithms which solve problems of the general class described by Eq.~\ref{eq:sparse_problem} is presented. In this presentation, a number of computational tools each node must be equipped with are described and example routines are provided for each. We conclude this presentation with a discussion emphasizing the consequences of several critical design decisions. Finally, in Section~3, an example algorithm adhering to the general framework is presented and evaluated.

	\subsection{Notation and nomenclature}
		Vectors will be denoted using an underscore and vector superscripts will be used to index vectors as opposed to indexing vector elements, i.e.~$\underline{x}^{(a)}_{1}$ and $\underline{x}^{(a)}_{2}$ represent distinct coordinates of the same vector. For simplicity of exposition we proceed assuming all vectors are in $\mathbb{R}^{N}$, extensions to complex-valued and/or $N_1\times N_2$-dimensional vector spaces follow readily. In order to avoid confusion we henceforth restrict the word {\it node} to only mean a graph vertex and reserve the use of the term {\it vertex} to that of a polytope.

\section{Unveiling the tree}
\label{sec:general_framework} 
	The general strategy underlying the framework described in this section is to convert the convex constraint satisfaction problem in Eq.~\ref{eq:sparse_problem} to a problem of graph traversal on a rooted tree. Every node then processes a set of elements belonging to $\mathcal{S}$, either received from initialization or a parent node, in order to produce a new set of elements which also belong to $\mathcal{S}$ but have fewer non-zero coordinates. Using this set of elements the node unveils if it has children of its own; if so it may pass its set of solutions to one or more of its children and the process repeats. The topology of the tree directly relates to the sparsity of feasible solutions in two ways: (i) the number of non-zero coordinates in the set of feasible solutions decreases with each successive generation of the tree and (ii) siblings of each group possess feasible elements with distinct sparsity patterns. 

	\subsection{Initializing the root node}
		We begin by equipping the root node with a set $\mathcal{P}$ of $M$ possibly distinct elements drawn from the feasible set $\mathcal{S}$ in Eq.~\ref{eq:sparse_problem}, i.e.
		\begin{equation}
			\mathcal{P} = \left\{\underline{x}^{(i)} \colon \underline{x}^{(i)}\in\mathcal{S}, \,1\leq i\leq M  \right\}
		\end{equation}
		where each $\underline{x}^{(i)}$ is generally dense. Note that $\mathcal{P}$ is the vertex representation of a particular polytope embedded within $\mathcal{S}$, i.e. the convex hull of the elements of $\mathcal{P}$ forms a closed convex polytope contained within $\mathcal{S}$. For problems where $\mathcal{S}$ is defined or naturally described using half-space representation, i.e. as a finite collection of linear equality and inequality constraints, standard vertex enumeration algorithms may be directly used to populate $\mathcal{P}$ \cite{vertexEnumeration}. Alternatively, the use of convex programs with random or systematically chosen objective functions can be used to obtain either well-spread vertices of $\mathcal{S}$ or elements with other desirable properties. As will become evident shortly, any greedy algorithm adhering to the proposed framework will not be able to obtain an optimal solution to Eq.~\ref{eq:sparse_problem} which is contained in $\mathcal{S}$ but not in the convex hull of $\mathcal{P}$\footnote{This discludes algorithms which successfully transform leaves into parents by drawing additional elements from $\mathcal{S}$, as described in Section~2.2.1.}.

		A pertinent question at this stage involes the selection of the design parameter $M$. To address this issue we refer to \cite{NumberOfVertices} in which a cyclic polytope in an $N$-dimensional space with $v$ vertices is shown to achieve the maximum number of obtainable facets $k$. Therefore, the corresponding dual polytope maximizes the number of vertices for a fixed number of facets where $v=\mathcal{O}(k^{\left\lfloor \frac{N}{2}\right\rfloor })$. This result renders complete vertex enumeration of $\mathcal{S}$ computationally intractable in practice without imposing additional structure. We defer further comments on computational complexity to Section~2.2.3.

	\subsection{Incremental sparsity: computation at each node}
		Each node of the tree, after having received a convex polytope $\mathcal{P}$ in vertex representation from its parent node (or initialization for the root), is equipped with three primary functions: (i) \textsc{unveilChildren}, (ii) \textsc{vanishCoordinate}, and (iii) \textsc{reduceComplexity}. In what follows we say that a node vanishes coordinate $d$ when the result of its processing produces a new polytope $\mathcal{P}'$ such that 
		\begin{equation}
			\underline{x}_{d}=0,\qquad \forall\underline{x}\in\mathcal{P}'.  \label{eq:vanish_d}
		\end{equation}
		Further, each node of the tree satisfies the following property: for every $\underline{x}\in\mathcal{P}',\,\underline{x}_{k}=0$ for all coordinates $k$ vanished along the unique path to the root. In addition, every node in generation $g$ contains a set of feasible solutions $\underline{x}$ to Eq.~\ref{eq:sparse_problem} which satisfy $\|\underline{x}\|_0 \leq N-g$.

	\subsubsection{\textsc{unveilChildren}}
		A fundamental ability required of each node in order to unveil the tree is to identify which, if any, offspring said node can produce. Stated another way, each node needs to be equipped with a subroutine which identifies the set of coordinates it can vanish given the polytope $\mathcal{P}$ received from its parent node.
		A sufficient condition for a particular node to vanish coordinate $d$ requires its received polytope $\mathcal{P}$ to contain at least one element $\underline{x}^{(+)}$ such that $\underline{x}^{(+)}_{d}>0$ and at least one element $\underline{x}^{(-)}$ such that $\underline{x}^{(-)}_{d}<0$. When this condition is satisfied, we refer to the potential offspring as child $d$. Let $\mathcal{I}$ denote the set of all such possible children for a given node, then that node may produce a maximum of $|\mathcal{I}|$ children. Pseudocode~\ref{funct:unveilChildren} describes the  subroutine \textsc{unveilChildren} which generates $\mathcal{I}$ for a given $\mathcal{P}$. A node is then classified as a leaf of the tree provided that it cannot generate any children, i.e. $|\mathcal{I}|=0$.

		Termination criteria for algorithms adhering to the framework proposed in this paper include, but are not limited to, identifying either the longest or a sufficiently long path from the root to a leaf. Subtree exploration, i.e. transforming a leaf node into a parent, is sometimes achievable by drawing additional elements from $\mathcal{S}$. For example, by drawing elements using a convex program where the objective function is designed to target the sign of a specific coordinate while simultaneously imposing additional constraints to vanish all coordinates along the direct path to the root.

		\begin{pseudocode}[t]
			\caption{A subroutine which determines the set $\mathcal{I}$ of potential children for a given set of feasible solutions $\mathcal{P}$.}\label{funct:unveilChildren}
			\begin{algorithmic}
				\Function{unveilChildren}{$\mathcal{P}$} \\
					$\hspace{1em}\mathcal{I} \gets \left\{d \colon \exists\, \underline{x}^{(+)},\underline{x}^{(-)}\in\mathcal{P} \mbox{ s.t. } \underline{x}^{(+)}_{d} > 0 \mbox{ and } \underline{x}^{(-)}_d < 0 \right\}$ 
				\EndFunction
			\end{algorithmic}
		\end{pseudocode}

		\begin{pseudocode}[t]
			\caption{A subroutine which populates $\mathcal{P}'$ using elements of $\mathcal{P}$ by vanishing coordinate $d$ using Eq.~\ref{eq:convexCombination} for $\ell = 2$. }\label{funct:embedPolytope}
			\begin{algorithmic}
				\Function{vanishCoordinate}{$\mathcal{P}, d$}
					\For{each $\underline{x}^{(+)}\in\mathcal{P}$ with $\underline{x}_{d}^{(+)} > 0$}
						\For{each $\underline{x}^{(-)}\in\mathcal{P}$ with $\underline{x}_{d}^{(-)} < 0$} \\
							$\hspace{4.2em}\mathcal{P}'\gets \left(\frac{\underline{x}_{d}^{(+)}}{\underline{x}_{d}^{(+)}-\underline{x}_{d}^{(-)}}\right)\underline{x}^{(+)} + \left(\frac{-\underline{x}_{d}^{(-)}}{\underline{x}_{d}^{(+)} - \underline{x}_{d}^{(-)}}\right)\underline{x}^{(-)}$
						\EndFor
					\EndFor
				\EndFunction
			\end{algorithmic}
		\end{pseudocode}

	\subsubsection{\textsc{vanishCoordinate}}
		Consider a node having received polytope $\mathcal{P}$ and fix some $d\in\mathcal{I}$ which we wish to vanish in the sense of Eq.~\ref{eq:vanish_d}. The general strategy by which the set $\mathcal{P}'$ is populated is presented next. In order to ensure that elements of $\mathcal{P}'$ maintain feasibility with respect to Eq.~\ref{eq:sparse_problem} and simultaneously satisfy Eq.~\ref{eq:vanish_d}, we make explicit use of the properties of a convex combination. Specifically, let $\underline{x}^{(1)},\cdots,\underline{x}^{(\ell)}\in\mathcal{P}$, then
		\begin{equation}
			\underline{x}' = \sum_{i=1}^{\ell}\alpha_i\underline{x}^{(i)} \in \mathcal{P} \hspace{1em} \text{for} \hspace{1em} \sum_{i=1}^\ell\alpha_i =1, \alpha_i\geq 0.\label{eq:convexCombination}
		\end{equation}
		It is straightforward to then show that given the qualification for $d$ to be an element of $\mathcal{I}$, there exists a set of linear combination weights $\alpha_{1},\dots,\alpha_{\ell}$ satisfying Eq.~\ref{eq:convexCombination} such that $\underline{x}'_{d}=0$. Any such combination for which this holds may then be used to systematically generate an element of $\mathcal{P}'$. 

		Pseudocode~\ref{funct:embedPolytope} contains the subroutine \textsc{vanishCoordinate} which generates this polytope using Eq.~\ref{eq:convexCombination} with $\ell = 2$ and a proper selection of the weights $\alpha_{1}$ and $\alpha_{2}$. In addition, Eq.~\ref{eq:convexCombination} implies both that the coordinates previously vanished in generating $\mathcal{P}$ remain vanished during the population of $\mathcal{P}'$ and that since $\underline{x}^{(+)}_d \neq \underline{x}^{(-)}_d \neq 0$ then $\mathcal{P}'\subseteq\mathcal{P}$, i.e. $\mathcal{P}'$ is embedded within $\mathcal{P}$.

		Depicted on the left of Figure~1 is two generations of a partially unveiled tree. Let $\mathcal{P},\mathcal{P}_{1}$, and $\mathcal{P}_{2}$ denote the polytopes on the highlighted path belonging to nodes ``root node'', $\underline{x}^{(6)}$ (child 6), and $\underline{x}^{(2)}$ (child 2), respectively. A cross-section of the polytope embedding  $\mathcal{P}_2 \subseteq \mathcal{P}_1 \subseteq \mathcal{P} \subseteq \mathcal{S}$ is depicted on the right. Also depicted is a different walk, i.e. ``root node'' to $\underline{x}^{(2)}$ (child 2) to $\underline{x}^{(6)}$ (child 6), ending with a set of vertices belonging to a different group of siblings demonstrating another set of elements with the same sparsity pattern as the elements of $\mathcal{P}_{2}$.

		\begin{figure}[t]
		  \centering
		  \centerline{\includegraphics[width=3.5in]{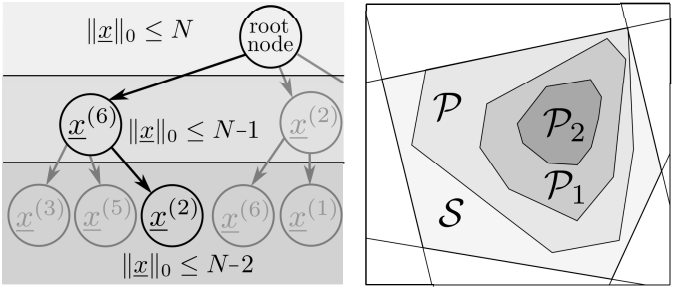}}\label{fig:NestedPolytopeFigure}
		  \caption{An example depicting two generations of a partially unveiled tree (left) and a cross-section of the polytope embeddings (right) corresponding to the highlighted walk from the root node to child 6 to child 2.} 
		\end{figure}

	\subsubsection{\textsc{reduceComplexity}}
		The computational complexity required to compute all possible elements of $\mathcal{P}'$ as described by the example routine \textsc{vanishCoordinate} in Pseudocode~2 is generally intractable, especially for problems with long root-to-leaf distances. Indeed, let $\mathcal{P}$ denote the received polytope of a node and let  $M^{(+)}_{d}>0$ denote the number of elements $\underline{x} \in \mathcal{P}$ satisfying $\underline{x}_{d} > 0$ and $M^{(-)}_{d}>0$ the number of elements satisfying $\underline{x}_d < 0$ for some $d\in\mathcal{I}$. Further, let $\mathcal{P}'$ denote the result of vanishing coordinate $d$ using \textsc{vanishCoordinate}. Then $\mathcal{P}'$ results in  
		\begin{equation}
			M'_d=M^{(+)}_{d}M^{(-)}_{d}
		\end{equation}
		possibly repeated vertices. It immediately follows that the total number of vertices generated for a given node in generation $g$ is on the order of $\mathcal{O}(M^{2^{g}})$ where the root nodes polytope was populated using $M$ vertices. By limiting the number of vertices used to generate an embedded polytope with coordinate $d$ vanished, the problem of tree traversal remains computationally tractable. This is explicitly at the expense of excluding regions of the polytope $\mathcal{P}$ in forming $\mathcal{P}'$ where the maximally sparse solution may reside. 

		A simple procedure to control the computational burden at each node is described in Pseudocode~\ref{funct:reduceComplexity} where the function \textsc{reduceComplexity} limits the number of vertices satisfying $\underline{x}_{d} > 0$ and $\underline{x}_{d} < 0$ to $\widehat{M}^{(+)}_{d} < M^{(+)}_{d}$ and $\widehat{M}^{(-)}_{d} < M^{(-)}_{d}$, respectively. In our experience with several examples, the number of unique vertices generated by \textsc{vanishCoordinate} is smaller than $M'_{d}$, especially in later generations of the tree. This heuristic may help guide the dynamic selection of $\widehat{M}^{(+)}_{d}$ and $\widehat{M}^{(-)}_{d}$ for a particular problem.

		\begin{pseudocode}[t]
			\caption{A function which reduces the computational complexity required to vanishing coordinate $d$ using \textsc{vanishCoordinate}.}\label{funct:reduceComplexity}
			\begin{algorithmic}
				\Function{reduceComplexity}{$\mathcal{P}, d$} \\
					$\hspace{1em}\mathcal{P}'\gets$ at most $\widehat{M}^{(+)}_{d}$ unique vertices from $\mathcal{P}$ with $\underline{x}_d > 0$ \\
					$\hspace{1em}\mathcal{P}'\gets$ at most $\widehat{M}^{(-)}_{d}$ unique vertices from $\mathcal{P}$ with $\underline{x}_d < 0$ 
				\EndFunction
			\end{algorithmic}
		\end{pseudocode}

		\begin{pseudocode}[t]
			\caption{An example of a single-path pre-walk coordinate selection rule for runtime execution.}
			\begin{algorithmic}
				\Function{selectCoordinate}{$\mathcal{P,\mathcal{I}}$} \\
					$\hspace{1em}d' \gets \displaystyle{\arg\max_{d\in\mathcal{I}}}|\{\underline{x}\in\mathcal{P}\colon \underline{x}_d > 0\}|\cdot|\{\underline{x}\in\mathcal{P} \colon \underline{x}_d < 0\}| $
				\EndFunction
			\end{algorithmic}
		\end{pseudocode}

	\subsection{Tree-search protocols}
		A number of instrumental design decisions informed by the problem at hand must be made when crafting a greedy algorithm of the general framework presented in this paper. For example, a critical decision involves selection of the tree traversal protocol which will be implemented, e.g., a depth-first or breadth-first search. This decision is obfuscated by the fact that the trees topology is unknown at the outset. When considering a breadth-first search, it is important to understand that while the width of the tree $w_{g}$ at generation $g$, i.e. the total number of nodes across generation $g$, is upper bounded by 
		\begin{equation}
			w_{g} = w_{g-1}(N-g + 1)\label{eq:widthBound}
		\end{equation}
		where $w_{0} = 1$, the details of the problem at hand may significantly restrict the width for a number of reasons resulting in breadth-first protocols which are much more computationally attractive than those implied by Eq.~\ref{eq:widthBound}.

		In considering depth-first protocols, the selection of either a run-time order selection algorithm or a predetermined order may depend critically upon the availability of heuristics or side information. For example, using the indices of a magnitude sorted $1$-norm relaxation of Eq.~\ref{eq:sparse_problem} or any other available side information may result in solutions of higher sparsity or earlier identification of deep leaves. Pseudocode~4 presents an example run-time elimination order subroutine \textsc{selectCoordinate} for a depth-first protocol which uses no heuristic or side information but generally aims to select coordinates to vanish which maximize the number of nodes in the embedded polytope $\mathcal{P}'$ produced using \textsc{vanishCoordinate}.

	\subsection{Subtree exploration}
		Techniques which are agnostic to the details of the particular problem at hand but attempt to restart the tree upon discovery of a leaf, or even more generally determine if there are children other than those identified using \textsc{unveilChildren} on a particular nodes polytope during runtime, can easily be incorporated into the design of an algorithm to allow further exploration of the trees topology. We refer to such a routine as a subtree exploration routine. One such technique, as mentioned previously in Section 2.1.1, is to attempt to transform leaves into parent nodes and depends directly upon the computational ease at which additional elements of $\mathcal{S}$ may be drawn with the addition of appropriately vanished coordinates. The class of algorithms as described until now is easily modified such that these types of subtree exploration attempts can be made at either the discovery of a leaf node or after no node in the unveiled tree has unexplored children.

		Another subtree exploration technique, which may be used to both trim branches and identify unexplored children, involves ending the exploration down a given node $a$ having received polytope $\mathcal{P}_{a}$ in a fixed generation $g$ for which another node $b$ in the same generation has received polytope $\mathcal{P}_{b}$ with elements consisting of the same sparsity pattern. Denote the union of the vectors in the two polytopes as $\mathcal{P}_{a\cup b}$, i.e.
		\begin{equation}
			\mathcal{P}_{a\cup b} = \left\{\underline{x} \mid \underline{x}\in\mathcal{P}_{a} \mbox{ or } \underline{x}\in\mathcal{P}_{b}\right\}
		\end{equation}
		and assign this polytope to node $b$. Then the algorithm continues from node $b$ and exploration is discontinued through node $a$. This technique is equivalent to searching for potential children nodes inside the convex hull of the two polytopes.

\section{Numerical example}
\label{sec:numerical_example}
	Maximally sparse filters, despite being a computationally difficult design problem, result in systems which have demonstrable advantages as compared to dense systems with respect to a number of practical metrics \cite{sparseFilterDesign}. Although sparse filter design formulated in half-space representation does meet the constraint qualifications of the compressive sensing framework, many well-known existing design methods have a similar flavor to those used in compressive sensing, i.e. they are broadly classified as greedy iterative algorithms which make use of, e.g.,  weighted $1$-norm linear programs \cite{dennis1}. A fair comparison with methods such as IHT and CoSaMP cannot be made in part due to various assumptions being violated. Further approaches, such as non-convex programs, have been used to produce filters with very sparse impulse responses \cite{dennis2}.  The use of alternative convexity principles has also been previously applied to filter design, e.g., in \cite{POCS2} an initial design is iteratively projected between two convex sets resulting in a final design satisfying the constraints of both sets but is not explicitly optimal with respect to any metric.
 
	In this section we generate an example algorithm adhering to the general framework proposed in Section~2 and apply it toward the design of a sparse, causal, Type I linear-phase finite-impulse-response lowpass filter $h[n]$ with support  $[0,2N]$.   In particular, let $\Omega_{pb}$ and $\Omega_{sb}$ denote the respective passband and stopband where 
	\begin{equation}
		\Omega_{pb}=\{\omega_{k} \colon \omega_{k}\in[-\omega_{pb},\omega_{pb}]\}\
	\end{equation}
	and
	\begin{equation}
		\Omega_{sb}=\{\omega_{k} \colon \omega_{k}\in[-\pi, -\omega_{sb}]\cup[\omega_{sb},\pi]\}
	\end{equation}
	for $0 < \omega_{pb} < \omega_{sb} < \pi$ and $1 \leq k \leq K$ where $K$ is chosen to sufficiently sample the frequency axis with respect to the filter support. We impose frequency-domain attenuation constraints such that a candidate impulse response is said to be feasible if its Fourier transform amplitude deviates no more than $\delta_{pb}$ and $\delta_{sb}$ from the ideal amplitude response over the passband and stopband, respectively. Let the ideal amplitude response, denoted by $D(\omega)$, be unity on $\Omega_{pb}$ and zero on $\Omega_{sb}$. Then, using the notation in Eq.~\ref{eq:sparse_problem}, the feasible set $\mathcal{S}$ is written as 
	\begin{eqnarray}
		\mathcal{S}=\{\underline{x}\in\mathbb{R}^{N} :&	\left|  T\left(\omega,\underline{x}\right)-D\left(\omega\right)\right|\leq\delta_{pb},	\omega\in\Omega_{pb},\hspace{0.5mm} \\
		&\left|T\left(\omega,\underline{x}\right)-D\left(\omega\right)\right|\leq\delta_{sb},	\omega\in\Omega_{sb}\} \nonumber
	\end{eqnarray}
	where 
	\begin{equation}
		T(\omega, \underline{x}) = \sum_{k=1}^{N}\underline{x}_k\cos\left(\omega \left(k - 1\right)\right)
	\end{equation}
	and $\underline{x}_1=h[0]$ and $\underline{x}_k=2h[k - 1]$ for $2\leq k \leq N$. The design example as formulated in this section is easily extended to describe other classes of filters possibly including additional constraints, e.g., in \cite{POCS} a number of common filter constraints are formulated as closed convex sets. The design specifications, similar to those found in an example in \cite{sparseFilterDesign}, are chosen as follows: passband cutoff frequency $\omega_{pb}=0.20\pi$, stopband cutoff frequency $\omega_{sb}=0.25\pi$, passband ripple $\delta_{pb}=0.01$ (linear), stopband ripple $\delta_{sb}=0.1$ (linear), and support parameter $N=31$. 

	Algorithm~1 describes a generic single-path depth-first algorithm using the example run-time order selection subroutine  \textsc{selectCoordinate} from Pseudocode~4. In order to apply Algorithm~1 to the sparse filter design problem, the root nodes initial polytope $\mathcal{P}$ is populated using $M=500$ vertices drawn from $\mathcal{S}$. The vertices are in particular selected by solving a sequence of linear programs where the coefficients of the objective vector are chosen uniformly in $[-1,1]$. In order to retain tractability, the subroutine \textsc{reduceComplexity} is used with $\widehat{M}^{(+)},\widehat{M}^{(-)} \leq 500$ at each generation, i.e. the polytope $\mathcal{P}$ cannot exceed $250,000$ vertices at any given generation. The algorithm, as described, does not attempt subtree exploration, i.e. no attempt is made to draw further samples in order to transform a leaf into a parent and thus the termination criterion is the discovery of a leaf. A randomly selected element of the leaf nodes polytope, transformed into an impulse response $h[n]$, is depicted in Figure~2. The length of the root-to-leaf walk for this example is $15$ and is directly related to the sparsity of the obtained impulse response.

	\begin{algorithm}[t]
	\caption{A single-path, depth-first algorithm with randomized vertex reduction and a run-time order selection subroutine.}
	$\hspace{1em}\mathcal{I} \gets \textsc{unveilChildren}(\mathcal{P})$
	\begin{algorithmic}
	\While{$\mathcal{I}\neq\emptyset$} \\
	$\hspace{1em}d \gets$  \textsc{selectCoordinate}($\mathcal{P},\mathcal{I}$)\\
	$\hspace{1em}\mathcal{P}\gets$  \textsc{reduceComplexity}($\mathcal{P}, d$)\\
	$\hspace{1em}\mathcal{P}\gets$ \textsc{vanishCoordinate}($\mathcal{P}, d$)\\
	$\hspace{1em} \mathcal{I} \gets \textsc{unveilChildren}(\mathcal{P})$ 
	\EndWhile
	\end{algorithmic}
	\end{algorithm}

	\begin{figure}[t]
	  \centering
	  \centerline{\includegraphics[width=3.6in]{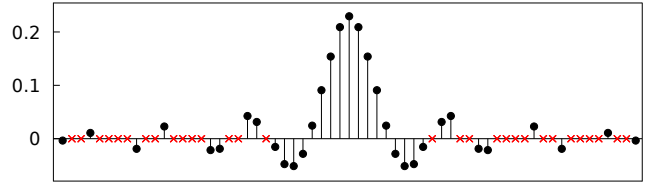}}\label{fig:filter}
	  \caption{An impulse response corresponding to one sparse solution generated using Algorithm~1. Zero valued coefficients are marked with red x's.} 
	\end{figure}

	Variations and extensions to Algorithm~1 follow immediately, such as utilizing alternative order selection rules as a function of the current sparsity level or pattern. For example, a natural alternative is to select a fixed ordering prior to runtime corresponding to the indices of the magnitude-sorted coefficients of the solution to the $1$-norm relaxation of Eq.~\ref{eq:sparse_problem}. In comparing these two path selection rules for a number of examples, the order selection rule in Pseudocode~4 tends to result in sparser impulse responses. This may in part be understood by the fact that thresholding the $1$-norm solution generally does not degrade gracefully from the frequency-domain constraints. Using subtree extensions and other variations generally resulted in impulse responses of different sparsity levels and patterns.

\bibliographystyle{IEEEbib}
\bibliography{refs}

\end{document}